\documentstyle[12pt,amssymbols]{article}
\input math_macros.tex

\begin{document}
\begin{titlepage}
\today          \hfill
\begin{center}
\hfill    LBL-37163rev \\

\vskip .5in

{\large \bf The Hard Problem: A Quantum Approach.}
\footnote{This work was supported by the Director, Office of Energy
Research, Office of High Energy and Nuclear Physics, Division of High
Energy Physics of the U.S. Department of Energy under Contract
DE-AC03-76SF00098.}

\vskip .25in
{\bf{Henry P. Stapp}}

{\em Theoretical Physics Group\\
    Lawrence Berkeley Laboratory\\
      University of California\\
    Berkeley, California 94720}
\end{center}

\vskip .5in

\medskip

\centerline{\bf{Contents}}

\begin{enumerate}
\item Introduction: Philosophical Setting.
\item Quantum Model of the Mind/Brain.
\item Person and Self.
\item Free-Will.
\item Qualia.
\item Meeting Baars's Criteria for Consciousness.
\end{enumerate}
\medskip
\medskip
Prepared for a special issue of the Journal of Consciousness Studies.
\end{titlepage}
\renewcommand{\thepage}{\roman{page}}
\setcounter{page}{2}
\mbox{ }

\vskip 1in

\begin{center}
{\bf Disclaimer}
\end{center}

\vskip .2in

\begin{scriptsize}
\begin{quotation}
This document was prepared as an account of work sponsored by the United
States Government. While this document is believed to contain correct
 information, neither the United States Government nor any agency
thereof, nor The Regents of the University of California, nor any of their
employees, makes any warranty, express or implied, or assumes any legal
liability or responsibility for the accuracy, completeness, or usefulness
of any information, apparatus, product, or process disclosed, or represents
that its use would not infringe privately owned rights.  Reference herein
to any specific commercial products process, or service by its trade name,
trademark, manufacturer, or otherwise, does not necessarily constitute or
imply its endorsement, recommendation, or favoring by the United States
Government or any agency thereof, or The Regents of the University of
California.  The views and opinions of authors expressed herein do not
necessarily state or reflect those of the United States Government or any
agency thereof or The Regents of the University of California and shall
not be used for advertising or product endorsement purposes.
\end{quotation}
\end{scriptsize}

\vskip 2in

\begin{center}
\begin{small}
{\it Lawrence Berkeley Laboratory is an equal opportunity employer.}
\end{small}
\end{center}

\newpage
\renewcommand{\thepage}{\arabic{page}}
\setcounter{page}{1}

\noindent{\bf{Introduction: Philosophical Setting }}

\medskip

In the keynote paper David Chalmers  has defined ``the hard problem'' to be the
problem of  integrating consciousness,  per se, into our  conception of nature.
``Consciousness'',  per se,  consists of {\it  experiences},  such as an actual
experience  of a  pain, or of a  sorrow, or  of a  redness. It  includes a {\it
visual experience}  of a table in a  room as distinguished  from an essentially
theoretical  construct, ``the table  itself'' that we  conceive, or imagine, or
believe to exist even when no one is experiencing it.

John Searle (1992) in his recent book ``The Rediscovery of the Mind'' has given
a  brief  account of  the  recent  history  of an   important   movement in the
philosophy of  mind, namely  materialism,  which tries to  evade the problem of
consciousness   by  denying  either  the  existence of    consciousness, or its
relevance  philosophy  and  science, or by  trying to  reduce  consciousness to
something  else,  for  example to   ``matter''---as matter  is  conceived of in
classical    mechanics---or to  some  functional  entity,  such as  the logical
structure of a computer  program. Searle gives  brief arguments, and cites more
detailed ones, which seem to show that  all materialist approaches tried so far
have failed, essentially because they do not include an essentially irreducible
component of reality, namely consciousness, to which he ascribes a first-person
or subjective mode of existence. This mode of beingness he distinguishes from a
third-person  or  objective mode of  existence,  which is the  mode ascribed by
classical mechanics to the particles and fields that constitute the irreducible
elements of that particular conceptualization of the world.

To explain this  notion of a  third-person, or objective,  mode of existence we
recall that classical  mechanics was created to  explain the motions of planets
and falling apples, etc. During early
childhood each of  us forms the  theoretical idea that  certain things, such as
his playthings,  exist  independently of their  being  experienced by himself
or
anyone else. Classical  mechanics is predicated  precisely on the related
notion that there
are,    similarly, tiny  invisible   objects  (particles),  and also unseen
wave-like structures  (fields), that are similar to  planets
in that they can be  conceived to exist  independently of anyone's experiences.
Thus an object, such as a human brain,  for example, is represented within this
idealized conception of  nature, classical  mechanics, as being {\it completely
made up} of these particles and fields that are supposed to exist independently
of anyone's experience.

The likely inadequacy of this simple idealization is, of course, manifest from
the outset.
An alert human brain is normally connected to someone's experience.
Thus there is no {\it \'a priori} reason to assume that we should be able to
adequately conceptualize this complex organ as merely a simple aggregation of
tiny localized entities that, like planets, can be imagined
to exist independently of anyone's experience.
Rather, one would naturally expect that certain properties of an actual brain
might become lost, or impossible to comprehend, within the framework of such an
idealization.

Searle's proposed solution of the problem of consciousness has three main
points:

{\ul{Point 1}}

``Consciousness is just an ordinary biological feature of the world'' (p. 85)

``The brain causes certain mental phenomena, such as conscious mental states,
and
these are simply higher-level features of the brain.'' (p. 14)

{\ul{Point 2}}

``Conscious mental states and processes have a special feature not possessed by
other natural phenomena, namely subjectivity.'' (p. 95)

``What more can we say about this subjective mode of existence? Well, first it
is
essential to see that in consequence of its subjectiveity, the pain is not
equally accessible to any observer.
Its existence, we might say, is an irreducibly first-person ontology'' (p. 95)

{\ul{{Point 3}}

``What I want to insist upon, ceaselessly, is that one can accept the obvious
facts of physics---for example that the world is made up entirely of particles
in fields of force---without in any way denying the obvious facts about our
existence---for example that we are  all conscious and that our conscious
states have quite specific irreducible phenomenological properties.'' (p. 28)

``One  can be a   thorough-going  materialist  and not in  any way  deny the
existence of   (subjective, internal,  intrinsic,  and often  conscious) mental
states.'' (p. 54)

Points 1 and 2 are plausible enough: consciousness could quite conceivably be
a natural property of the brain that is `higher-level',  in the sense that it
is left  out of the classical idealization of the brain, and hence is not
reducible to the third-person ontology that characterizes classical mechanics.

Point 3 is also  plausible, to the  extent that one does  not try to comprehend
the particles, fields, and matter of Searle's thorough-going materialism as the
classical-mechanics   idealizations of  these  things. For these  idealizations
have, by  virtue  of the way  in which  they are   conceived of and  defined in
classical   mechanics, a  purely  third-person  beingness.  The  causal laws of
classical    mechanics can  cause  these   particles and   fields, as  they are
conceptualized in  classical mechanics, to coalesce  into all sorts of causally
efficacious  functional entities, but  nothing within  those classical laws, as
they are conceived of in  classical mechanics, can cause the emergence of
some ``new mode of  beingness'' that goes beyond  the beingness of aggregates
of particles and fields.  This is because classical  mechanics is a theory that
was based, from the outset, on the idea that everything is nothing more than an
aggregations  of things  that have  only  third-person  beingness: first-person
beingness was explicitly excluded at the outset, and all causal connections are
explained  within classical  mechanics in  terms of aggregates  of third-person
things acting in concert. Since all functional entities constructed in this way
are {\it  causally}  reducible to  third-person  entities there  is no rational
place in the theory for the re-introduction of first-person beingness.

The conclusion that ought to be drawn  from Searle's conclusion---which is that
there are two  different modes  of beingness,  with  first-person beingness not
reducible to third-person beingness, but constituting, nevertheless, a natural
feature  of  organs such  as  brains---is  that the   idealizations  upon which
classical mechanics was  based are not adequate to  describe such organs: a new
kind of  mechanics is needed;  one that  naturally ascribes two different
modes of beingness to such organs.

This conclusion drawn from Searle's philosophic analysis might seem at first to
conflict with science. Indeed, the motivation of the materialists was evidently
to bring philosophy into  accord with science,  which in the nineteenth century
meant  classical mechanics,  with its monistic  ontology. But  we now know that
classical  mechanics fails to  describe  correctly the  properties of materials
such as, for  example, the  tissues of a human  brain.  Classical mechanics has
been superceded by quantum mechanics, which is characterized, above all, by the
fact that it is dualistic: the single  monistic ontology of classical mechanics
is replaced by an  ontology consisting  two very different  kinds of beingness.
One kind of beingness is  the kind enjoyed by the  quantum mechanical analog of
the   ``matter'' of  classical   mechanics.  This part  of  nature,  namely the
deterministically   evolving  wave function,  is like  the matter  of classical
mechanics in the  sense that it is  represented as an  aggregation of localized
properties,   and  the  temporal   evolution of  each of  these   properties is
determined exclusively by neighboring  properties, in accordance with equations
of motion that are direct  analogs of the  corresponding equations of classical
mechanics.   However, this  in only  half the of  the  quantum  story: there is
necessarily a  second component of the  quantum ontology,  one that pertains
to {\it choices} between alternative possible {\it experiences}.

Searle,  when,  confronted  by the  suggestion  that  quantum  theory, with its
inherent dualistic  ontology, is important to the  resolution of the mind-brain
problem, says  that he will  wait until  quantum theorists  come into agreement
among themselves  about the  interpretation of the theory.  But that misses the
point   completely. All   interpretations  agree on  the  need for a  dualistic
ontology, with  one aspect  being the quantum  analog of {\it  matter}, and the
other aspect  pertaining to {\it  experiences}. Thus the whole debate among
quantum theorists  is essentially a  debate about  the mind-matter connection.
This debate is  precisely where  an input from philosophy of mind should enter.
To wait until  the quantum  debate is over is  to miss the whole mind-matter
ball game.

This point is important enough to elaborate upon, at least briefly.
I shall therefore describe here the five main approaches to quantum theory,
focussing on the dualistic and mind-versus-matter aspects of each.

The most orthodox  of the  interpretations of quantum  theory is the Copenhagen
interpretation, as  expressed in the  words of Niels  Bohr. The key idea is
encapsulated in two quotations:

``In our description of nature the  purpose is not to disclose the real essence
of phenomena but  only to track down  as far as possible  relations between the
multifold aspect of our experience'' [Bohr, 1934]

``Strictly    speaking, the    mathematical  formalism  of  quantum  theory and
electro-dynamics    merely  offers rules of   calculation for the  deduction of
expectations pertaining to observations  obtained under well-defined conditions
specified by classical physical concepts.'' [Bohr, 1958]

Bohr is emphasizing here that science,  in the end, has to do with correlations
among our  experiences:  experiences  are the  ultimate data  that science must
explain.  Thus he can  renounce the  classical  ideal of giving  a mathematical
description of  the objective  world itself in  favor of  constructing a set of
mathematical rules that allow us to  compute expectations pertaining to certain
kinds of experiences. Thus, in  contrast to the the ideas of classical physics,
human  experiences have an essential  place in the theory. Yet the mathematical
formulation of the  ``rules of  calculation'' is based on  a description of the
``matter-like'' aspect mentioned above.

This approach is  dualistic because the two things  that it deals with are,  on
the one  hand,  our {\it   experiences }  (of a  certain  special  type, namely
classically    describable   perceptions)  and,  on the  other  hand,  a set of
mathematical rules  that allow us to  compute expectations  pertaining to these
experiences, and these rules  are  expressed in terms  of a  generalization of
the mathematical structure that occurred in classical  mechanics, and that
represented, in that idealization, the ``objective world of particles and
fields''.

Bohr's pragmatic approach was  revolutionary in its day, and was firmly opposed
by most of the senior scientists of that time. In Einstein's opinion:

``Physics is an attempt conceptually to grasp reality as it is thought
independently of its being observed'' [Einstein, 1951, p.81]

 and quantum theory, as formulated by Bohr,

``offers no useful point of departure for future developments''
[Einstein, 1951, p.87]

Bohr   admitted, in  fact,  that his  form of  the  theory  would not  work for
biological  systems. That, of course,  was the origin of a  logical gap between
the two  parts of his  orthodox  formulation of the  theory,  i.e., between the
subjective   (experiential)  part  associated  with our brains, and the
objective (material) part associated with the experiments that human
scientists perform on atomic systems.

Under the pressure of  diverse goals (e.g., to  expand the scope of the theory
to include biological and cosmological systems,
or to firm up the logical foundations) a number of ``ontological formulations''
of quantum  theory have  been created.  They  attempt to give a  picture of the
entire world itself, not just a set of rules that allow us to form expectations
about our future experiences.

The simplest ontology is that of David  Bohm (1952). In the orthodox (Bohr)
theory one spoke of the complementary ``particlelike''  and ``wavelike''
aspects of a quantum  system. That was  confusing  because  particles stay
confined to tiny regions while waves spread  out: the two  concepts contradict
each other, physically. This is what forced Bohr into his epistemological
stance, and his idea of ``complementarity''.

For a world consisting of a single  quantum entity Bohm's model would have both
{\it a  particle} and  {\it a wave}:  the particle  rides like a  surfer on the
wave. One easily sees how  the puzzling double-slit  experiment is explained by
this model: the wave goes  through both slits and  influences the motion of the
particle,  which goes  through just  one slit. This  model is  dualistic in the
sense of  having both a  particle and a wave.  But this  dualism is basically a
mind/matter   dualism,   because the  function  of the   ``particle'',  or more
specifically its generalization to the  many-particle universe, is basically to
specify what  {\it our  experiences}  will be.  There is a huge  gap in quantum
theory between the  information  contained in the ``wave''  and the information
contained in our  experience. The  purpose of, and need  for, the particle, and
its generalization  to the  many-particle universe, is  basically to supply the
information---not contained in the  wave (function)---that specifies {\it which
one} of the many  mutually  incompatible experiences  allowed by quantum theory
the  observer   actually  has. If  there  were no  need  to  describe  the {\it
experiential}  aspects of reality,  which are very  different in character from
what the  deterministically evolving wave  (function) describes, there would be
no need for the  ``particle-part'' of Bohm's  ontology. The critical assumption
in Bohm's  model is  precisely  the  assumption that  even though  the ``wave''
(i.e., wave  function of the universe)  might describe a  superposition of many
different brains of some one  particular scientist, say Joe Smith, and although
each these different superposed ``brains'' would correspond to his perceiving a
different result  of some experiment  that he is  performing, only one of these
brains will  actually be  illuminated by the  light of  consciousness, and this
particular brain,  the one that  possesses consciousness,  is picked out by the
``particle'' aspect of the theory, in a specified mechanical way.

To explain how  this (and also  the other  models) work, I  shall often use the
term ``branches of the wave  function''. To visualize these branches, imagine a
large pond with an  initially smooth  surface (no waves).  A source of waves is
placed at the center, but is  surrounded by a barrier that has some gaps. These
gaps allow  ripples to spread  out only along  certain  beam-like regions, with
most  of the  surface  of the  pond  remaining   smooth. These  well  separated
beam-like regions of propagating  ripples I call ``branches'', or ``branches of
the wave (function)''.

The surface of a pond is only two  dimensional. But the quantum-mechanical wave
that corresponds to a universe consisting of $N$ particles would be a wave in a
$3N$-dimensional space. The ``branches  of the wave (function)'' will typically
be relatively  narrow beams of waves  in this  $3N$-dimensional space, and each
beam will  correspond, in a typical  measurement  situation, to some particular
``classically  describable'' result of  the measurement.  For example, one beam
may  describe,  at some  late  stage, a  particle  detector  having  detected a
particle;  {\it  and} a  corresponding  pointer  having  swung to  the right to
indicate that the detector has detected the particle; {\it and} the eye and the
low-level processing parts of the brain responding to the light signal from the
pointer in  the  swung-to-the-right  position; {\it  and} the  top-level neural
activity  that  corresponds to  the  observer's  perceiving the
pointer in the swung-to-the-right  position: the other branch would describe
the
particle   detector's having  failed  to detect  the  particle; {\it  and } the
pointer  remaining in  the  center position;  {\it and}  the eye  and low-level
processing parts of the  brain responding to the  light signals coming from the
pointer  in the  center  position;  {\it  and} the   top-level neural  activity
corresponding to  the observer's  perceiving the pointer in the
center  position. The  fact that {\it  both}  branches of the  wave are present
simultaneously is not  surprizing once one  recognizes that the wave represents
essentially only a {\it probability for an experience to occur}: there is, in a
typical  measurement, a possibility  for each of several  possible experiential
results to  occur, and the  probability  function (or wave  function) will then
have a ``branch'' corresponding to each possibility.

  Of   course,  the   observer,  Joe  Smith,  will  see  only  one  of  the two
possibilities: he will see {\it either} the pointer swung-to-the-right {\it or}
the or  the  pointer  remaining at the  center  position.  To  accommodate this
empirical fact  Bohm introduces his  ``surfer'' in the  $3N$-dimensional space.
The surfer is merely a point in the  $3N$-dimensional space that move always in
a direction  defined by  the shape of  the  $3N$-dimensional  wave at the place
where this point  is, and this rule of  motion for the  surfer ensures that the
surfer will end up in one  branch or another, not  in the intervening ``still''
part of  the  $3N$-dimensional  space. Each  branch  corresponds  to one of the
possible experiences. If the  ``surfer'' (which is just the moving point in the
$3N$-dimensional    space)  ends  up in  the  branch  that   corresponds to the
experience ``I see the pointer in the swung-right position'' then, according to
Bohm's theory,  this  perception of the  pointer  ``swung-to-the-right'' is the
experience  that actually  occurs: only the  single branch in  which the surfer
ends up will be  ``illuminated''; all others ``remains  dark''. Bohm's rules
for the  motion of  the  surfer ensure  that if  the  various  possible initial
conditions for the surfer are assigned appropriate ``statistical weights'' then
the statistical predictions of his theory about what observers will experiences
will agree  with the  those  given by the  orthodox  (Bohr) rules.  In this way
Bohm's  deterministic model  reproduces the quantum statistical  predictions
about what our experiences will be.

The two parts of Bohm's  ontology, namely the wave  in the 3N-dimensional space
and the `surfer', can  both be considered  `material', yet they are essentially
different because the waves describe  all the possibilities for what our actual
experiences   might be,  and  therefore  has a  beingness  that is  essentially
``potential'', whereas the trajectory of the surfer specifies the actual choice
from among the various alternative possibilities, and therefore has a beingness
that  represents  ``actuality''  rather than  mere   ``potentiality'': the wave
generates all the {\it possible} experiences, whereas the trajectory defined by
the surfer specifies which one of these possible experiences actually occurs.

Bohm's   model is  very  useful,  but as  a  model of  reality  it has  several
unattractive features.  The first is the ``empty  branches'': once two branches
separate they generally move further  and further apart in the $3N$-dimensional
space,  and  hence  if the   ``surfer''  gets in  one  branch  then  all of the
alternative ones become  completely irrelevant to  the evolution of experience:
the huge set of empty branches  continues to evolve for all of  eternity, but
has no effect upon anyone's experience.

A more  parsimonious  ontological  theory, not  having these  superfluous empty
branches,  was  described by  Heisenberg  (1958). It  also  involves a  reality
consisting  of two  kinds of  things. His two  kinds of  things are   ``actual
events'', and ``objective tendencies for those events to occur''. The objective
tendencies can be taken  to be represented by the  wave on the 3$N$-dimensional
pond, and the actual  events can be represented by  sudden or abrupt changes in
this wave. Each such change ``collapses the wave'' to one of its branches. Thus
Bohm's ``surfer'', which specifies a {\it choice} between branches, is replaced
by an ``actual  event'', which  also specifies  a choice  between branches. But
whereas Bohm's  surfer has no  back-reaction on the wave,  each of Heisenberg's
actual   events   obliterates  all  branches  but  one.  The big   problem with
Heisenberg's  theory is to  find a reasonable  criterion for  the occurrence of
these actual events.

Wigner  (1961) and  von Neumann  (1932),  noting that  there is  nothing in the
purely  material  aspect of  nature that  singles out  where the  actual events
occur, suggest that these events should occur at the points where consciousness
enters:   i.e., in    conjunction with   conscious   events.  This is  the most
parsimonious possibility: all of the  known valid predictions of quantum theory
can be reproduced by limiting the actual events to brain events that correspond
to experiential  events. An argument  based on survival of  the species [Stapp,
1995a]  provides support  for the idea  that actual  events  occurring in human
brains will tend  to occur at the  brain-wide level of activity that
corresponds  to   conscious  events,  rather than  at some   microscopic (e.g.,
molecular, or individual-neuron) level.

This Wigner-von-Neumann version of Heisenberg's theory will be discussed
presently in some detail. But first a few remarks about the final major
interpretation are needed.

In the  Everett  many-minds  theory the  basic quantum  mechanical  equation of
motion,  the   Schroedinger  equation,  holds  uniformly:  there are  no sudden
collapses of the wave function; all branches continue to exist. Moreover, it is
assumed  that,  because all of  the branches  exist,  all of the  corresponding
streams of conscious must also occur.

Since  the  various   branches  propagate  into   different  parts of  the $3N$
dimensional space  they will evolve  independently of each  other: the physical
``memory   banks'',   associated with  one  branch  will not  effect  the brain
activities  specified by  another branch. Hence each different  branches can be
considered to define different  ``self'', or ``psyche'', with  each of
these selves  continually dividing  into different extentions of itself
into the future.

At first sight this idea seems to  allow the whole theory to be reduced to just
one entity, the evolving  wave, with the different  psychological persons being
just ``aspects'' of  corresponding brain activities  on different branches. But
that is not  correct. The  branches of the  wave function  appear as parts of a
{\it conjunction} of branches: all branches on the `pond' exist simultaneously,
even though they  evolve  independently. But the  predictions of quantum theory
are an essential part of the theory,  and these statistical predictions pertain
to experiences that are `this experience' {\it or} `that experience', not `this
experience' {\it  and} `that  experience'. To speak of  probabilities one needs
something with an {\it or} character: something that can become associated with
{\it    either}  this   branch  {\it  or}  that   branch,  not  both
simultaneously.  Just as  the  different branches  of the wave  on the pond are
conjunctively present and do not, by  themselves, provide any ontological basis
for assigning different probabilities to these simultaneously present things,
so also is the quantum wave by itself insufficient for this task.

In Bohm's  theory  this extra  element of  the theory  was the  `surfer', which
determined the  experiences of the observers; in  Heisenberg's theory the extra
things were  the actual  events, which also  determined the  experiences of the
observers. In the Everett  interpretation the only  existing things besides the
waves are our  experiences, and there  is supposed to be a  separate experience
for with  each branch.  Thus we end up  again with  a dualistic  theory; with a
world that is composed of the one ``material'' universe represented by the wave
function,  which evolves  always  according to the  the  Schroedinger equation,
plus, for each  named person, an great  profusion of many  minds, or streams of
consciousness:  the stream of  consciousness  of Joe Smith  must be continually
splitting into different  separate branches, with  at least one for each of the
perceptibly different results of any experiment that he performs. Consequently,
the  proponents  of the  theory  need to  develop,  in order  to  complete this
interpretation, some coherent dualistic ontology  involving, for each of us, a
profusion of  branching  minds, each known only to itself.

In summary,  all the major  ontological  interpretations of  quantum theory are
dualistic,  in the sense  that they  have one  aspect or  component that can be
naturally  identified as the  quantum analog  of the {\it  matter} of classical
mechanics, and a second aspect that is associated with {\it choices} from among
the  possible  {\it  experiences}.  All   interpretations  are, in  this sense,
basically similar to the Wigner-von-Neumann interpretation to be explored here,
but are less parsimonious, in that  they involve either existing but unobserved
branches  (Bohm), or  existing but  unobserved  actual events  (Heisenberg), or
existing but unobserved  minds (Everett).  (Parenthetically I note that the one
great   virtue of  the   Everett    interpretation,  namely  that   requires no
faster-than-light    influences, will  probably  evaporate  when the  theory is
completed by the construction of the  needed consistent theory of the observing
minds, for the theory will then run up against the nonlocality  theorems: See
Mermin, 1994)

I now return to  philosophy---from this digression  pertaining to the dualistic
character of  quantum  theory---and comment  briefly upon one  of the principal
contemporary  versions of   materialism, namely  `eliminative  materialism', as
expounded in the recent book {\it  Neurophilosophy} by P.S. Churchland (1986) .
There it is noted  that there are  familiar examples in  the history of physics
where a theory dealing with one realm  of phenomena, for example thermodynamics
or optics, has been  reduced to a `more basic'  theory, for example statistical
mechanics or  electrodynamics. So why cannot  psychology be likewise reduced to
brain physiology, and ultimately to the basic physics of matter? Searle answers
that in all of these  familar reductions the  psychological part of the problem
was ``carved  off'' before the  reduction was  achieved, so  the analogy is not
apt: no new  kind of  beingness has  ever been  obtained from  the third-person
beingness of classical physics. Churchland avoids this ontological issue of the
nature or quality of the beingness by restricting the notion of reducibility to
the {\it causal}  properties of the theories in  question, thereby skirting the
issue that Searle focusses upon. However, she must eventually face the issue in
the form of the problem of explaining the seemingly huge difference between, on
the one hand,  things such as pains,  desires, beliefs,  and other experiential
things, and, on the other hand, material particles. She deals with this problem
by suggesting that the psychology of  the future may be very different from the
`folk    psychology' of  today:  it  may not   contain  such  things as  pains,
perceptions, and other experiential  things. However, as Chalmers emphasizes in
his keynote  paper, that  kind of  `solution'  eliminates the  very facts to be
explained  by   psychological  theory, and  in  fact, as  stressed by  Bohr, by
physical theory as well. As Searle  maintains, an adequate theory of the future
ought to represent experiences as natural features of biological organs, rather
than explaining them away.

Achieving    Searle's    desideratum  requires  no  waiting  for  some  unknown
theory-of-the-future,  for quantum  theory, combined with  some rather standard
ideas from  neuro-science,  already presents  us with an ideal  psycho-physical
framework.

\noindent{\bf{2. Quantum Model of the
Mind/Brain}}

\medskip

The main  features of the  mind/brain theory  proposed in  Stapp (1993) are now
briefly described.

1. {\ul{Facilitation}}:  The pattern of  neurological  activity associated with
   any occurring  conscious thought is  ``facilitated'', in  the sense that the
   activation  of this  pattern causes  certain  physical changes  in the brain
   structure,  and  these changes  {\it facilitate}   subsequent  activations
of this
   pattern.

2. {\ul{Associative Recall}}: The facilitation of patterns mentioned above is
   such that the excitation of a part of a facilitated pattern has a tendency
to
   excite the whole.
   Thus the sight of an ear tends to activate the pattern of brain activity
   associated with a previously seen face of which this ear was a part.

3. {\ul{Body-World Schema}}: The physical body of the person in its
   environment is represented within the brain by certain patterns of neural
   and other brain activity. Each such pattern has {\it components}, which are
   sub-patterns that represent various parts or aspects of the body and its
   environment, and these components are normally patterns of brain activity
   that have been facilitated in conjunction with earlier experiences.

4. {\ul{Executive-Level  Template for Action:}} A  main task of the alert brain
   at each moment is  to construct a  template for the  impending action of the
   organism. This template is formed from patterns of neural and brain activity
   that,  taken  together,   represent a   coordinated plan  of  action for the
   organism. This   representation is  implemented by the  brain by means of an
   automatic causal spreading of neural excitations from the executive level to
   the rest of  the nervous  system.  This  subsequent activity  of the nervous
   system causes both motor responses and lower-level neural responses.

The   executive-level  templates are  based on  the  body-world  schema, in the
following  sense.  There are two  kinds of  templated  actions:  attentions and
intentions.  Attentions  {\it up date}  the  body-world schema:  they bring the
brain's  representation of the body in  its environment up  to date. Intentions
are  formulated in  terms of a  {\it  projected} (into  the  future) body-world
schema:  they  are  expressed  in  terms of an  image  of how  the  body in its
environment is  intended to be at a  slightly future time.  (Thus, for example,
the tennis player imagines how he will strike the ball, or where the ball he is
about to hit will land in his opponent's court).

5. {\ul{Beliefs and other Generalizations}}: The simple Body-World Schema, with
   attentional  and intentional  templated  actions, is the  primitive level of
   brain action: it gives the general  format. However ``beliefs'' can be added
   to the  landscape.  Also, each  templated  action has  both  intentional and
   attentional aspects.

6. {\ul{Quantum Theory}}: The features mentioned above are key elements of this
   theory.  But they  are aspects  that hold  at the  level  corresponding to a
   particular  classically  described  `branch'.  However,  classical mechanics
   cannot  account for the  properties  of the  materials (such  as tissues and
   membranes)  from which the  brain is made.  Hence, within  the basic theory,
   these  classically  describable  aspects  must be  coherently  imbedded in a
   correct quantum mechanical description if one is to have an adequate account
   of the behavior of the brain.

7. {\ul{Superposition   of  Templates}}: An  analysis  [Stapp,  1993, 1995b] of
   processes   occurring  in  synapses  shows  that if  there  were no  quantum
   collapses occurring in brains then a brain evolving according to the quantum
   laws must evolve, in general, into a  state that contains a superposition of
   different ``branches'', with each of  these branches specifying the template
   for a different  macroscopic action:  each of these  different templates for
   action will  evolve into  a different  response  of the  nervous system, and
   consequently into a difference macroscopic response of the organism.

8. {\ul{The Reduction Postulate}}: Following the Wigner-von-Neumann approach, I
   postulate that the  quantum collapse of the brain  state occurs at the level
   of the template for action: the (Heisenberg-picture) state (of the universe)
   undergoes the collapse  $$ \Psi_i\to\Psi_{i+1} =  P_i \Psi_i, $$ where $P_i$
   is a  projection  operator that acts  on  appropriate  macroscopic variables
   associated with the  brain: it picks out and  saves, or ``actualizes'', {\it
   one} of the  alternative possible  templates for action,  and eradicates the
   others.  Hence the  organism will  then proceed  automatically  to evolve in
   accordance  with this one  particular plan  of action,  rather than evolving
   (\'a la Everett) into a  superposition of states  corresponding to {\it all}
   of the different possible  macroscopically distinguishable courses of action
   that  were   formerly  available  to it.  Thus  the   ``quantum  event'', or
   ``collapse of the wave function'', {\it selects} or {\it chooses} one of the
   alternative possible coherent plans  of action---previously generated by the
   purely    mechanical    functioning  of  the   brain---by    actualizing the
   executive-level pattern of brain  activity that constitutes {\it one} of the
   alternative possible templates for action.

This collapse of  the wave function is  to be understood  not as some anomalous
failure of the laws of nature, but  rather as a natural consequence of the fact
that wave  function does not  represent  actuality itself, but  rather, in line
with the ideas of Heisenberg, the  ``objective tendencies'' for the next actual
event.

Each such event is represented, within the Hilbert space description, as a
sudden
shift in the wave function, or state $\Psi_i$, to a new form that incorporates
the conditions or requirements imposed by the new actual event.

These   collapse  events  in  the  Hilbert  space  are not   things  introduced
willy-nilly: they are needed to block  what will otherwise automatically occur,
namely the evolution of  the wave function to a  form that directly contradicts
collective  human  experience:  all of us  who see the  pointer  agree that the
pointer does  not both  swing to the  right {\it  and} also  remain motionless.
Under the conditions of  the measurement it does  one thing {\it or} the other,
and all of us who witness  what it does, and are  able communicate our findings
to each  other, agree  about which one  of these  two possible  things actually
occurs.

9. {\ul{The Basic Postulate}}: Adhering to the Wigner-von-Neumann approach, I
   postulate that this physical brain event, namely the collapse of the wave
   function to the branch that specifies one particular template for action, is
   the brain correlate of a corresponding psychological or experiential event.
   Thus the psychological experience of ``intending to raise the arm''
   corresponds to the physical event that actualizes the template for action
   that ``tends to raise the arm''. The pschological event of ``intending to do
   $x$'' is paired to the physical event that ``tends to do $x$.''

Attending is a special kind intending: the intention, in the case of attending,
is to up-date the body-world schema.

Different  locutions can be used here.  One can say that  the brain event is an
image in the  world of  matter of the  conscious  event, or that  the conscious
event  is the  image in  the world  of  mind of the  brain  event, or  that the
conscious  event and  brain event are  two aspects  of one and  the same actual
event. But the  essential point is  that the  quantum-mechanical description of
nature  in   terms of  the     deterministically   evolving  wave   function is
fundamentally    incomplete: some   ontological  element  that is  structurally
different and distinct from the local-deterministically evolving wave function,
which represents  the quantum analog  of matter, is needed  to specify the {\it
choices  between  alternative  possible  experiences}.  This added  ontological
element is  {\it  logically needed} in  order to  provide a  basis for the core
property   of   quantum   mechanics,  namely  its   capacity  to   predict {\it
probabilities for classically describable experiences to occur}.

10. {\ul{The Efficacy of  Consciousness}}: In this model the choices associated
    with conscious events are dynamically efficacious: each such event {\it
    effects} a decision between different templates for action, and these
    different templates for action lead to different distingushable responses
of
    the organism.

11. {\ul{Consciousness and  Survival}}: It is often  claimed that consciousness
    comes into being because it aids  survival. For this to be so consciousness
    must be  efficacious. Yet (just as in  classical physics)  consciousness is
    not   efficacious in the  Bohm  and  Everett  models: everything is
    completely  pre-determined.  Consciousness would be  nonefficacious also in
    the  Heisenberg  model if we did not follow   Wigner-von-Neumann in
    associating (at least some of) the actualizing events with conscious
events.

I am  {\it not}  assuming  that {\it  all}  actual events  are  associated with
physical events in human  brains: other events may  also occur. The assumption,
rather, is that every conscious event is efficacious and hence corresponds to a
physical  event. One must  expect, in an  organism whose  physical structure is
determined  in  large  measure by   considerations related  to  survival of the
species, that these  physical events will in fact  occur primarily at the level
of the   actualizations of the  top-level  templates  for action,  because this
placement provides the optimal survival advantage.[Stapp, 1995a, 1995b]

12. {\ul{Conscious  Events and  Unconscious Processing}}:  The general temporal
    development  in the  brain proceeds  by periods  of  unconscious processing
    punctuated by conscious events. A conscious event actualizes a template for
    action that, by the automatic spreading of top-level neural activity to the
    rest of the  nervous system,  controls motor action,  the collection of new
    information   (including  the  monitoring of  ongoing   processes), and the
    formation of the next template for action.

Classically  only a  single ``next  template''  would be formed.  This could be
achieved  either by the  formation of a  resonant state that  sucks energy from
competing possibilities, or by inhibitory signals, or by dropping into the well
of an  attractor. But in  any case the  quantum  uncertainties  entail that the
quantum  brain  will   necessarily  evolve  into a   superposition of  branches
corresponding  to the  different alternative  possible  classical templates for
action. Next the quantum event in the  brain selects one of these templates for
action, and then the automatic  (unconscious) neural processes proceed to carry
out the  instructions  encoded  in the  template. Thus  we have an  alternation
between   discrete   conscious   events---each of  which   decides  between the
alternative  possible allowed  templates for action  generated by the automatic
action of the local  deterministic laws of quantum  mecanics, and hence between
the different associated macroscopic responses of the organism---and periods of
unconscious activity controlled by the local deterministic laws.

\newpage
\noindent{\bf{3. Person and Self}}

\medskip

According to William James:

``Such a  discrete   composition is  what  actually  obtains in our  perceptual
experience. We either perceive nothing, or something that is already there in a
sensible  amount. This fact is  what is known  in psychology  as the law of the
`threshold'. Either your experience is of no content, of no change, or it is of
a perceptible amount of content or change. Your acquaintance with reality grows
literally by buds or drops of  perception. Intellectually and on reflection you
can divide these into components, but as immediately given they come totally or
not at all.''[James, 1910, p. 1062]

``... however complex the object may be  the thought of it is one undivided
state
of consciousness.''[James, 1890, p. 276]

``The consciousness of Self involves a stream of thought, each
part of which as  `I' can (1)  remember those that went  before, and know the
things they knew; and (2) emphasize and care paramountly for certain ones among
them as {\it `me'}, and {\it  appropriate} to these the rest... This {\it me}
is an empirical aggregate  of things objectively  known. The {\it I~} that
knows
them cannot itself be an aggregate.  Neither for psychological purposes need it
be  considered to be  an  unchanging  metaphysical  entity like  the Soul, or a
principle like the pure Ego,  viewed as ``out of time''. It is a {\it Thought},
at each moment different from that of  the last moment, but  {\it appropriative
of the  latter},  together with  all that  the latter  called its  own ... {\it
thought is itself the thinker}, and psychology need not look beyond...''
(James. 1890, p. 401)

In line  with these  ideas of James,  and those of  the  preceding section, the
conception of a `person' that  emerges here is that of a sequence of discrete
psychological  (i.e.,  experiential or  conscious) events   bound together by a
matter-like structure, namely the  brain/body, which evolves in accordance with
the local  deterministic laws of  quantum mechanics. Each  conscious event is a
new entity  that rises  from the  `ashes' of the  old, which  consists of the
propensities for its occurrence carried by the brain/body.

A felt sense of  an enduring  `self'  is  experienced, and hence it
must, within this  theory, be  explained as an  aspect of  the structure of the
{\it  individual}  discrete  conscious  events. The  explanation  is this: each
conscious event has a ``fringe'' that surrounds the central image, and provides
the background in which the central image is placed. The slowly changing fringe
contains the consciousness of the  situation  within which the immediate action
is taking place; the historical setting including  purposes (e.g., getting some
food to  eat). The sense  of feeling  of self is in  this  fringe. It is not an
illusion,   because  the  physical  brain/body  is  providing   continuity and
a
reservoir of  memories  that can be  called upon,  even though  each thought
is,
according to this model, a separate  entity. As explained by James---see also
Stapp (1993)---each thought, though  itself a single entity, has components
that
are sequentially  ordered in a  psychological time, and  hence each thought has
within its own  structure  an aspect  that corresponds to  the flow of physical
time.
\newpage

\noindent{\bf 4. Free-Will}

\medskip

Among the qualia that we experience is  the feeling that we are, in some sense,
free.  That  is an   accurate  feeling.  The  whole  organism  is free  to make
high-level  choices. Its  fate is not   predetermined, and its  actions are not
controlled by mechanical local deterministic laws in a way that would make that
feeling of freedom a complete illusion.

It might be objected that we are not free because, according to quantum theory,
our choices are determined by blind chance. That misses the point. In the first
place the choices are not blind. If the quantum events in the brain occurred at
the level of the neurons then the  choices would be blind, for the consequences
of each   individual choice  would be  screened  from view  by the  inscrutable
outcomes of  billions of  similar independent  random choices.  But the choices
being made by  the organism,  acting as a  unit, are choices  between plans for
actions that  have clear  and  distinctive  consequences for the  organism as a
whole, in terms of its future  behavior. The choice is made at the level of the
organism as a  whole, and the  event has a  distinctive `feel'  that accurately
portrays its consequences for the organism as a whole. The conditioning for
this event is an expression of the the  values and goals of the whole organism,
and the choice is implemented by a unified action of the whole organism that is
normally meaningful in  the life of the organism. And this meaning
is felt as an essential aspect of the act of choosing.

The  final  `random'  decision between  the  alternative  possible  distinctive
actions of the organism is not some  wild haphazard stab in the dark, unrelated
to the  needs  or  goals of  the  organism. It  is a  choice  that is  governed
essentially  by  the number  of ways  in which  the  mechanistic  aspect of the
organism,  which has been  honed to construct  templates for  action concordant
with the needs of  the organism within  its environment,  can come up with that
particular template. Thus the choice  is not like the throw of an unconditioned
die. It  is a  carefully  crafted  choice that  tends to  be the  ``optiminally
reasonable''  choice under the  conditions defined by the  external inputs, and
the needs and goals of the organism. Each of the alternative possible templates
for a coherent  and  well-coordinated action  of the organism  emerges from the
quantum soup, and is given, by the quantum mechanism, a weighting that reflects
the interests of the organism as a  whole, within the context in which he finds
himself. The  choice is  conditioned  by these  personally  molded weights, and
therefore tends to be a decision that is optimally reasonable from the point of
view of  the  organism.  This  arrangement  avoids  both the  Scylla  of a fate
ordained  and   sealed at  the  birth of  the  universe  by a   microscopically
controlled blind  mechanism, and also the Charybdis  of a haphazard wild chance
that operates at a microscopic level,  and is therefore blind as regards likely
consequences, and their  evaluations from the  perspective of the organism. The
intricate interplay of  chance and determinism  instituted by quantum mechanics
effectively  frees the  organism to  pursue, in an  optimal way,  its own goals
based on its own values,  which have themselves  been created, from a wealth of
open possibilities, by its own earlier  actions. Each human being, though never
in full control of  the situation in  which he finds  himself, does create both
himself and  his actions,  through a  process of a   microscopically controlled
deterministic   evolution  punctuated by  organic  meaningful  choices that are
top-down in  the sense that  each one is  instituted by an  actualization event
that  selects as a  unit,  and feels  as a unit,  an  entire  top-level plan of
action.

[Within the contemporary framework of quantum theory that I am adhering to here
there remains, in the end, an element  of `pure chance' that selects one of the
templates for  action `randomly'.  Whether this occurrence  of pure chance is a
permanent feature of basic physical theory, or merely a temporary excursion, no
one knows. In my  own opinion this  occurrence pure chance  a reflection of our
state  of  ignorance   regarding the  true  cause,  which must  in any  case be
nonlocal, and hence both  difficult to study and  quite unlike the local causes
that science  has dealt with  up until now. In  another place  [Stapp, 1995b] I
have described in more detail the  technicalities of the actualization process,
and also the possibility of replacing  the element of pure chance by a nonlocal
causal  process that  makes the  felt  psychological  subjective  `I', as it is
represented within the  quantum-theoretic description, rather than pure chance,
the source of the  decisions between one's  alternative possible courses of
action.]

\newpage
\noindent{\bf 5. Experience/Consciousness}

\medskip

The `Hard Problem'  is the problem of  {\it conscious  experience}: What is it?
Why is it present at all? Why is it so different from the other part of Nature,
namely the objective aspect of  reality? Why is it personal, or subjective? Why
is it  so  fleeting,  whereas  matter is  permanent  and  conserved?  Can it be
`reduced'  to  matter?  Can any  purely  physical  account  explain  it? Is the
material  of which  the brain is  made  crucial, or is  it only  the functional
aspect that is  critical? Why  is it so  closely connected to  function? How do
functional aspects  become ontological  aspects, i.e. how  does function become
being? How  can  anything, and  in  particular  consciousness, be  added to the
already closed laws of physics? Is  experience a fundamental element of nature,
or derivative, or emergent? What are  the {\it bridging laws} that connect mind
to matter?

Chalmers asks these questions, and says that right now we have no candidate
theory that answers these questions. But we do!

Chalmers  suggests that  perhaps there is a  small loop-hole  in quantum theory
that might provide an opening for  consciousness. But there is not just a small
loop-hole: there is a gigantic lacuna, which consists of fully {\it half of the
theory}, and this  hole provides an  ideal home for  consciousness. For quantum
dynamics consists  not only of the  mechanical process  that is governed by the
Schroedinger  equation, which  controls the  matter-like  aspect of nature, but
also an entirely  different `second process', which  constitutes a beingness of
an entirely different order. This  second process fixes the actual experiential
aspect of  nature, as  contrasted to  the potential  aspect. It  fixes what our
experiences  actually  will be. And in  the most  parsimonius of  the available
interpretations it {\it consists} of actualizations of precisely the functional
states that  we ``feel''  are being  actualized by  our  intentional mood. This
second  process is,  in  comparison to the  ontological   structures upon which
classical  mechanics  was based,  something  completely  new and  different: it
actualizes  things that  formerly were  mere  potentialities,  and hence has an
ontological status that  is different from the  ontological status of matter in
both classical  mechanics and  quantum  mechanics. It is a  `doer', and what it
does is just what our thought do, or  at least feel that they do: it initiates
physical and mental actions.  As an initiator of body/brain  action it is
indistinguishable from a stream of efficacious conscious thoughts.

How does this theory answer the questions raised above about consciousness?

What  is  consciousness?  It is a   sequence of   actualizations of  functional
patterns of brain activity. These functional patterns are expressed in terms of
a projected body-world schema, and each actualized pattern is `facilitated' for
use in later executive events.

Why are these actualizations present at all? Because the laws of physics demand
it. Without  such actualizations (or, in other interpretations, some substitute
for  them)  quantum  theory would  be  devoid of  empirical   significance, and
essentially  incomplete.  The  actualizations are  not  epiphenomenal! They are
efficacious,   and  hence can  play an  important role  in the  survival of the
organism.

Why is  consciousness so  different from the  other part of  Nature, namely the
objective   aspect of reality?  The objective  part of reality  has a different
kind of beingness:  it is mere  `potentia', whereas  consciousness is a doer;
it is a process of actualization.

Why  is   consciousness   subjective? It  is an   actualization  that  has many
components that are all integral parts  of the whole. The totality contains the
fringe of the experience that constitutes the `I', or `psyche', that is felt as
the experiencing  subject and  actualizer. The  experiencing subject is part of
the thought:  if it were  not part of  the thought  then there  would be in the
thought no  awareness of `I' as the  background relative  to which the focus of
the thought is the  foreground. So it not that the  thought belongs to the `I',
but rather that the `I' belongs to the thought.

Why is the  thought so  fleeting,  whereas matter  is permanent  and conserved?
Because a thought  is event-like,  whereas matter is the  continuously evolving
potentia for an event to occur.

Can consciousness  be `reduced' to  matter? ``Matter'' is  mere potentia for an
event. But each conscious event is  represented within matter (i.e., within the
wave) as  the  collapse of  the wave  (function)  to a  form that  embodies the
actualized functional structure. The  actualization cannot be expressed outside
of the matter that  embodies it, yet, by virtue of  its being an actualization,
it is not a mere potentia for such an actualization

Can any purely physical account  explain it? If by a physical account one means
a quantum mechanical account then the  actualization is an integral part of the
physical account, and is thus  explained by that account. But it cannot be
explained within the ontology of classical mechanics. For classical mechanics
has no events that are actualizations of potentia, and no concept of a potentia
that is a mere objective tendency for an actualization to occur.

Is  the  material  of which  the  brain  is made   crucial, or  is it  only the
functional  aspect  that is  critical? The  material  must support  the quantum
theoretic generation of the possible templates, and the actualization of one of
them. The conscious  process is a real process of  quantum actualization, not a
simulation  of that  process in  which this actualization does not actually
occur.

Why is consciousness so  closely connected to  function? In the specific theory
described here this close  connection arises  because the conscious event is an
actualization of a template for  action. The biological reason for this link of
actualization to function  is undoubtedly the  survival advantage it confers: a
species  constructed so that the  actualizations create  functionally effective
and reinforced actions will fare  better than one in which the created patterns
lack functional content.

How do functional  aspects become  ontological aspects?  Actualizations endow
structures with beingness.  Conscious actualizations  in  human brains endow
functional structures with material beingness: i.e., with the capacity to
tend to make certain later actualizations occur.

How can consciousness be  added to the already  closed laws of physics? Nothing
efficacious could be  added if the laws were  already complete! But the quantum
laws are grossly incomplete before consciousness, or some stand-in, is added.

Is  experience  a  fundamental   element of  nature,  or is it   derivative, or
emergent? An actualizing element that converts potentia to actuality is needed
to complete  quantum  theory. A  coherent role for  experience  is also needed.
Quantum theory allows these two needs to be satisfied together.

What are the {\it bridging laws} that connect mind to matter?
These laws have been described here, and in more detail in my book.

\noindent {\bf 6. Meeting Baars's Criteria for Consciousness.}

\medskip

Baars (1995) has formulated a set of empirical constraints that any sensible
theory of consciousness has to fit.

The first thing  that the theory must  account for is the  fact that there is a
great deal of unconscious processing  that is akin to consciousness, but is not
conscious. For example, there are  below-threshold and masked stimulations that
seem to be being processed in ways  akin to our conscious processing, but which
do not rise to  consciousness.

As described in Stapp (1993), the key units in brain processing are patterns of
excitations that have  been previously facilitated  and are called ``symbols''.
The task  of the  brain is  to  assemble some  subset of  these  symbols into a
coherent pattern of brain activity  that constitutes a coordinated template for
action. This  template is  expressed in a  `body-world  schema', which is the
brain's     representation of  the      body-in-its-environment,  or a  natural
generalization of this schema.

In the process of forming the next  template for action the input stimuli begin
to excite various  symbols. But a  great deal of automatic  (i.e., unconscious)
processing occurs before  there emerges from the  welter of competing symbols a
single coherent  combination of them that fits  together into single
coordinated
body-world schema. The  symbols activated by weak  stimuli, can {\it influence}
this competitive process  of creating the next  template, without these symbols
becoming   actually  represented  in the  final  template  itself:  they become
squeezed out by the requirement that the actualized template must form a single
coherent  body-world schema. This  picture of the general  mode of operation of
the unconscious process  of constructing the next  template seems to provide an
adequate basis (though, of course, not  the specific details) for understanding
the effects of weak or masked stimulations that Baars cites.

Perceptual processes are  understood in the same  way: the various symbols that
have been  activated all feed  into a  (quantum) mechanical  brain process that
must extract from this  welter of symbols, each of  which tends to excite other
symbols, a coordinated  combination of them that  fit together to form a single
coherent    body-world  schema,  before  any  conscious  event  can  occur. The
collection of  inputs excite  symbols that act  as a set of  clues from which a
single coherent schema must be formed. The fading from consciousness of stimuli
that call for no attentional or intentional action is accounted for by the fact
that the conscious events  correspond exactly to  events that either up-date or
project the  body-world schema, or  some natural  generalization of it. Symbols
that lack  the energy, or the relevancy  as defined  by the whole  active mass
of
competing symbols, to be  included in a current  template for action will not
be
experienced.

Why are   unaccessed  interpretations  of  ambiguous   interpretations not also
present in consciousness? The reason  is that an up-dating takes the form of an
actualization  of a coherent  body-world  schema. A coherent  body-world schema
must have definite  qualities assigned  to various points  in a spacetime grid;
all  ambiguities  must be  resolved  before the   body-world schema  comes into
being.. One  can surmise  that a  coherent  body-world schema  has the internal
dynamical     self-consistency  that  allows  it to   persist  long  enough for
facilitation  to occur.

Why is processing slowed down when two  alternative interpretations are closely
balanced in  likehood?  The reason is  that the  various stimuli excite the
associated symbols  and these patterns
tend to expand  to fill out  the body-world  schema. But if  there are balanced
tendencies  coming  from two   incompatible  alternatives  then the  mechanical
process requires more  time in order resolve the  conflict and produce a single
coherent body-world schema.

Another set of constraints mentioned by Baars are the contextual constraints on
perceptions. Again, in the process of constructing the next template for action
all the stimuli tend to  produce their  corresponding symbols (patterns). These
various  symbols  all enter  into the   unconscious  process of  constructing a
template for  action that  fills the  requirements  of being a  single coherent
body-world   schema.   Expectations, and  the needs  of the   organism, are all
represented  by input  symbols, and  this  collection of symbols constitutes an
initial  set  of  competing  patterns  that  must be  resolved  by the  brain's
automatic    machinery.  This   machinery   must, if  the   organism is  to act
effectively, create an appropriate template for a coordinated action that meets
the pressures (i.e., tendencies) that  are represented in the various initially
excited symbols.

Another  category of questions  raised by  Baars concerns not  percepts but
{\it
images}, for example the  visual images that we can  bring to mid when our eyes
are closed.

Where is our image of yesterday's breakfast before we bring it to mind?
Answer: In the patterns of activity that were facilitated yesterday at
breakfast, and hence exist as symbols that can be activated by the excitation
of some of its components, but that are not currently excited.

Why after a brief  exposure to a  visual matrix can we  access more information
than we can  report? Answer: because  the symbols  associated with the parts of
the matrix are all present in our  low-level brain response, but the processing
of this  information that  leads to an  up-dating  of the  body-world schema is
conditioned by the  ``need'' of the organism as  defined by other input stimuli
and the ``mental set''  defined by the preceding  conscious events, which issue
the instructions that are directing the construction of the next template. Only
a small  part of  the  welter of  input  symbols  makes it  through  the filter
provided by  the symbols  that  represent the  current  contextual situation to
become parts of the next template for action.

I can go through  the list given by  Baars and show that  all of his conditions
can be met, {\it at this level of  general principle}---as distinguished from a
description of specific  mechanisms at the neuronal  level---by using the ideas
used above. More  generally, this quantum picture  of the mind/brain seems {\it
compatible},  at this level of  general  principle, with all  of the mind/brain
data that I have  encountered in my perusal of the  literature. This perusal is
not  exhaustive, but  I think  covers enough  data to  make it  likely that the
general ideas described here will  adequately comprehend, at this general level
of description, what is  now known. Of course,  working out a detailed neuronal
machinery that will  implement these general  notions is a huge problem. But in
approaching that huge  problem it should be helpful  to have a rational general
conception  of  how  things  could  work in a  theory  of the   mind/brain that
encompasses in a coherent  way the fact that  classical physics does not give a
correct account of the behaviour of the materials out of which brains are made.

\newpage
\noindent{\bf

References}

Baars, Bernard (1995), ` A Thoroughly Empirical Approach to Consciousness',{\it
PSYCHE:   an     interdisciplinary  journal  of   research on    consciousness}
1039-723X/93.

Bohm, David (1952), `A Suggested Interpretation of the Quantum Theory in Terms
of Hidden Variables' I and II, {\it Phys. Rev. } {\bf 85,} pp. 166-93

Bohr, Niels (1934), {\it  Atomic Physics and Human  Knowledge}, Cambridge Univ.
Press, Cambridge, p. 18.

Bohr,  Niels   (1958), {\it   Essays  1958/1962  on  Atomic  Physics  and Human
Knowledge}, (Wiley, New York), 1963, p. 60.

Churchland, Patricia S. (1986), {\it Neurophilosophy}, (MIT Press, Cambridge).

Einstein, Albert (1951), in {\it Albert Einstein: Philosopher-Scientist}, P.~A.
Schilpp (ed.), (Tudor, New York).

Everett   III,  Hugh   (1957),  `Relative   State  Formulation of  Quantum
Mechanics', {\it Rev. of Mod. Phys. } {\bf 29}, pp. 454-62

Heisenberg, Werner  (1958), {\it  Physics and Philosophy}, (Harper and Row, New
York) Chap. III.

James, William  (1890), {\it The Principles  of Psychology}, (Dover, New
York, 1950) Vol. 1.

James, William (1910), {\it William James: Writings  1902-1910}, (The Library
of
America,   Viking, New  York,  1987), p. 1061. (From:  Some  Problems of
Philosophy, Ch X).

N. David Mermin (1994), `Quantum Mysteries Refined', {\it American Journal of
Physics} {\bf 62}, 880-887.

Searle, John (1992), {\it The Rediscovery of the Mind}, (MIT Press, Cambridge).

Stapp,   Henry P. (1993),  {\it  Mind,   Matter,  and   Quantum   Mechanics},
(Springer-Verlag, Heidelberg, Berlin, New York). Chapter 6

Stapp, Henry P. (1990), `Quantum Measurement and the Mind-Brain Connection',
{\it Symposium on the Foundations of Modern Physics 1990}, P. Lahti and P.
Mitterstaedt eds., (World Scientific, Singapore, 1991).

Stapp, Henry P. (1995a), `Quantum Mechanical Coherence, Resonance, and Mind',
To appear in the Proceedings of the Norbert Wiener Centenary Congress
(Edited by V. Mandrekar and P.R. Masini) to be published in the American
Mathematical Society Series Proceedings of Symposia in Applied Mathematics
(PSAPM) Also to appear in PSYCHE. Lawrence Berkeley Laboratory Report
LBL-36915.

Stapp, Henry P. (1995b) `Chance, Choice, and Consciousness: The  Role Of Mind
in the Quantum Brain. (For Tucson II and Psyche)

von Neumann, John (1932), {\it  Mathematical Foundations of Quantum Mechanics},
(Princeton University Press, Princeton, 1955}, Chap. VI.

Wigner, Eugene  (1961), `Remarks on  the Mind-Body  Problem' {\it The Scientist
Speculates}, I.J. Good, ed., (Heineman, London)

\end{document}